%% file: main.tex
\documentclass[acmsmall]{acmart}

\acmJournal{TECS}

\AtBeginDocument{%
  }

\usepackage{amsmath}
\usepackage{graphicx}
\usepackage{textcomp}
\usepackage{xcolor}
\usepackage{algorithmic}
\usepackage{array}
\usepackage[linesnumbered,ruled]{algorithm2e}
\usepackage[binary-units,per-mode=symbol]{siunitx}
\usepackage{textgreek}
\usepackage{pdflscape}
\usepackage{threeparttable} 
\usepackage{booktabs}
\usepackage{multirow}
\usepackage{float}
\usepackage{placeins}
\usepackage{color}
\usepackage[normalem]{ulem}
\usepackage{wrapfig}
\usepackage{subcaption}
\usepackage{flushend}
\usepackage{mdwlist}
\usepackage{makecell}
\usepackage{tabularx}     
\usepackage{adjustbox}    
\usepackage{url}
\usepackage{pifont}

\newcommand{\schedname}{MEDEA}

\newcommand{\cmod}[1]{#1}
\newenvironment{cmodb}{}{}

\begin{document}


\title{{\schedname}: A Design-Time Multi-Objective Manager for Energy-Efficient DNN Inference on Heterogeneous Ultra-Low Power Platforms}

\author{Hossein Taji}
\email{hossein.taji@epfl.ch}
\orcid{0009-0006-7794-5960}
\affiliation{%
  \institution{Embedded Systems Laboratory (ESL), EPFL}
  \city{Lausanne}
  \country{Switzerland}
}

\author{José Miranda}
\email{jose.miranda@upm.es}
\orcid{0000-0002-7275-4616}
\affiliation{%
  \institution{Industrial Electronics and Multimodal Systems Center (CEIMM), UPM}
  \city{Madrid}
  \country{Spain}
}

\author{Miguel Peón-Quirós}
\email{miguel.peon@epfl.ch}
\orcid{0000-0002-5760-090X}
\affiliation{%
  \institution{EcoCloud, EPFL}
  \city{Lausanne}
  \country{Switzerland}
}

\author{David Atienza}
\email{david.atienza@epfl.ch}
\orcid{0000-0001-9536-4947}
\affiliation{%
  \institution{Embedded Systems Laboratory (ESL), EPFL}
  \city{Lausanne}
  \country{Switzerland}
}

\renewcommand{\shortauthors}{Taji et al.}

\begin{abstract}
\input{secs/abs}

\end{abstract}



\maketitle

\section{Introduction} \label{sec:intro} \input{secs/intro}
\section{Related Work} \label{sec:rel_work}\input{secs/rel_work} 
\section{{\schedname}, Our Proposed Design-Time Multi-Objective Manager}\label{sec:manager} \input{secs/sched}

\section{Experimental Setup} \label{sec:setup} \input{secs/setup}

\section{Results and Analysis} \label{sec:results} \input{secs/results}

\section{\cmod{Limitations and Future Work}} \label{sec:limitations} \input{secs/limitations}
\section{Conclusion} \label{sec:conclusion} \input{secs/conclusion} 
\begin{acks}
\input{secs/acknowledgments}
\end{acks}


\bibliographystyle{ACM-Reference-Format}
\bibliography{mybib}

\end{document}

%% file: secs/abs.tex
The growing demand for on-device AI necessitates energy-efficient execution of DNN based applications on resource-constrained ultra-low power (ULP) platforms.
Heterogeneous architectures, combining specialized processing elements (PEs), have emerged as a key solution for achieving the required performance and energy efficiency.
However, optimizing energy while executing applications on these platforms requires efficiently managing platform resources like PEs, power features, and memory footprint, all while adhering to critical application deadlines.
This paper presents {\schedname}, a novel design-time multi-objective manager for energy-efficient DNN {inference} on Heterogeneous ULP (HULP) platforms.
{\schedname} uniquely integrates: kernel-level dynamic voltage and frequency scaling (DVFS) for dynamic energy adaptation; kernel-level {granularity scheduling,} suitable for specialized accelerators;
memory-aware adaptive tiling to navigate severe memory constraints; and all within a timing constraint-based optimization strategy, {which minimizes} energy based on application deadline.
To showcase practical viability, we evaluate {\schedname} on HEEPTimize, a heterogeneous ULP platform (22 nm, FPGA-prototyped) featuring a
RISC-V processor besides Near-Memory Computing (NMC) and Coarse-Grained Reconfigurable Array (CGRA) accelerators.
Experimental results, using a biomedical seizure detection case study, demonstrate that {\schedname} achieves overall energy reductions of up to $\SI{38}{\percent}$
compared to representative state-of-the-art methods, while consistently meeting all timing and memory requirements.
This effectiveness is attributed to its integrated features, with our analysis showing that kernel-level DVFS alone can be responsible for over $\SI{31}{\percent}$ of the energy savings in specific scenarios.

%% file: secs/intro.tex
The increasing prevalence of AI, particularly DNN, is driving a shift towards the deployment of AI processing at the edge, closer to the data source.
This paradigm offers compelling advantages over traditional cloud-based AI, including reduced latency, enhanced data privacy, lower bandwidth consumption, and offline operations capabilities \cite{ray2022review, schizas2022tinyml, scherer2024hardware}.
However, executing computationally demanding DNNs on resource-constrained edge devices presents significant challenges, especially with respect to power and memory. 
To address these, \textit{heterogeneous architectures} have emerged as a key solution \cite{machetti2024x}. 
By combining {heterogeneous} PEs (e.g., CPUs, specialized accelerators), architectural specialization allows significant improvements in both performance and energy efficiency compared to traditional homogeneous systems \cite{ponzina2022hardware}.

However, merely incorporating heterogeneous PEs is insufficient; Realizing the performance and energy benefits of heterogeneity depends on \textit{ effective resource management and scheduling} \cite{kang2020scheduling}. 
This involves intelligently assigning DNN computational tasks to the most suitable PEs \cite{risso2025optimizing, kakolyris2023road, kang2020scheduling, dagli2022axonn}, dynamically adjusting operating points (such as voltage and frequency) through DVFS \cite{liu2021dynamic, zidar2024dynamic},
and efficiently handling data movement \cite{burrello2021dory, stahl2023fused, jung2024optimizing}, while adhering to the application timing requirements \cite{kang2020scheduling, xiang2019pipelined} and the limitations of the platform resources \cite{ray2022review, schizas2022tinyml, scherer2024hardware}.
The challenge of resource management and scheduling becomes particularly acute in the \textit{ULP} domain, where platforms operate under exceptionally stringent constraints, often characterized by \si{\milli\watt} or \si{\micro\watt} power budgets and only \si{\kibi\byte}  to \si{\mebi\byte} of 
available on-chip memory \cite{ray2022review, schizas2022tinyml, scherer2024hardware}.
These severe limitations sharply distinguish ULP devices from the broader edge category, which includes significantly more 
capable platforms like mobile SoCs (e.g., \cite{kang2020scheduling, wang2020neural, karatzas2023omniboost}) or embedded systems such as the 
NVIDIA Jetson series (e.g., \cite{bouzidi2023map, odema2023magnas, kakolyris2023road, dagli2022axonn}).

Particularly relevant to the ULP domain is the increasing adoption of the RISC-V Instruction Set Architecture (ISA), preferred for its open, modular, and extensible nature in resource-constrained embedded systems \cite{machetti2024x}. 
This trend has led to several heterogeneous ULP (HULP) platforms based on RISC-V explicitly designed for energy efficiency, such as DIANA~\cite{ueyoshi2022diana}, DARKSIDE~\cite{garofalo2022darkside}, and HEEPocrates~\cite{machetti2024heepocrates}.
These platforms often integrate specialized hardware accelerators, crucial components of their heterogeneous design. Key paradigms include NMC units \cite{caon2024scalable}, 
which minimize costly data movement by computing near or within memory, and CGRAs \cite{denkinger2023acceleration}, which offer a balance of hardware efficiency and software flexibility for diverse DNN kernels. 
These, among other specialized digital accelerators \cite{garofalo2022darkside}, are vital to efficiently execute demanding applications such as DNN-based ones on ULP platforms. 
%

\begin{figure}[htbp]
    \centering
    \includegraphics[width=1\linewidth]{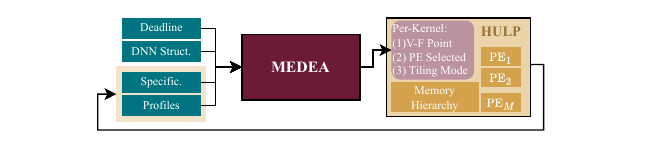} 
    \caption{Overview of the {\schedname} design-time multi-objective manager, illustrating its kernel-level decisions on PE, V-F, {and} Tiling mode, used to orchestrate energy-efficient DNN inference on a {HULP}.} 
    \Description{High-level MEDEA flow from a DNN workload and platform profiles to per-kernel processing-element, voltage-frequency, and tiling decisions for execution on a heterogeneous ultra-low-power platform.}
    \label{fig:sched_conceptual_overview}
\end{figure}

However, effective management and scheduling of DNNs on HULP platforms presents a unique confluence of challenges. 
Existing solutions
\cite{hamdi2024match, kakolyris2023road, liu2024adaknife, kang2020scheduling, burrello2021dory, stahl2023fused, risso2025optimizing, wang2020neural,
vasiliadis2022best, dagli2022axonn, bouzidi2023map, jeong2021deep, xiang2019pipelined, dagli2024shared, karatzas2023omniboost, odema2023magnas}
often address certain aspects of edge deployment, but struggle to holistically tackle the combined demands inherent to the ULP domain.
{In fact, the main} shortcomings, hindering optimal performance, include a lack of kernel-level DVFS, insufficient memory-aware adaptive tiling mechanisms for severe memory constraints, energy optimization frequently decoupled from strict application timing deadlines, 
coarse scheduling granularities are ill-suited for specialized accelerators, and often limited support beyond specific DNN architectures. 
To overcome these limitations, we propose \textbf{\schedname} (\textbf{M}anager for \textbf{E}nergy-efficient \textbf{D}NNs on h\textbf{E}terogeneous ULP \textbf{A}rchitectures), a novel design-time multi-objective manager specifically designed for HULP platforms.
\cmod{In this work, {\schedname} targets the dominant HULP deployment contract of periodic, single-model inference: one sequentially ordered workload $\mathcal{W}=\{k_1,\ldots,k_{N_{\mathcal{W}}}\}$ with a single application deadline~$T_d$.
Kernels execute strictly in the compile-time list order---the optimizer selects PE, V--F level, and tiling per kernel. 
The ILP/MCKP formulation runs offline on a host during deployment preparation; no solver is invoked on the HULP at inference time.}
Figure~\ref{fig:sched_conceptual_overview} presents an overview of {\schedname}, illustrating how, based on DNN description, timing constraint, and platform characterization data, 
it generates per-kernel scheduling decisions determining the optimal PE, V-F setting, and tiling strategy for efficient execution on {a HULP}.

The main contributions of this paper are as follows:
\begin{itemize}
    \item We introduce {\schedname}, a novel design-time multi-objective manager tailored for HULPs. 
    To the best of our knowledge, it is the first to integrate the following four features within a single, unified manager for HULPs: 
    (1) kernel-level DVFS for dynamic energy adaptation; 
    (2) kernel-level scheduling granularity suitable for specialized accelerators; 
    (3) memory-aware adaptive tiling to navigate severe memory constraints; all within a 
    (4) timing constraint-based optimization strategy that minimizes energy based on application deadlines. 

    \item  {\cmod{We demonstrate the real-world applicability and effectiveness of {\schedname} on periodic single-model inference through a Transformer-based biomedical seizure detection application (TSD).}
    Our evaluation includes a study on the impact of varying application deadlines on energy performance trade-offs and scheduling decisions made by {\schedname}. 
    This application study shows that our approach achieves energy reductions of up to $\SI{38}{\percent}$ compared to relevant baselines, while consistently meeting all timing requirements.}
           
    \item {We provide an in-depth analysis that isolates and quantifies the distinct energy savings enabled by {\schedname}'s core features. 
    Our evaluations demonstrate that, depending on the operational scenario and deadline, its kernel-level DVFS enables up to $\SI{31.3}{\percent}$ energy reduction, 
    memory-aware adaptive tiling up to $\SI{8.5}{\percent}$, and kernel-level scheduling up to $\SI{2.8}{\percent}$, 
    highlighting the individual importance of each integrated mechanism.}

\end{itemize}

The remainder of this paper is organized as follows:
Section~\ref{sec:rel_work} reviews state-of-the-art approaches on scheduling and energy optimization of DNN-based applications on heterogeneous platforms. 
Our proposed multi-objective manager, {\schedname}, is detailed in Section~\ref{sec:manager}.
Section~\ref{sec:setup} describes the overall experimental setup. 
Section~\ref{sec:results} presents and discusses the experimental results.
\cmod{Section~\ref{sec:limitations} states {\schedname}'s scope, modeling assumptions, and future extensions.}
Finally, Section~\ref{sec:conclusion} summarizes the main conclusions of this work.

%% file: secs/rel_work.tex
\begin{table*}[htbp] 
  \centering
  \caption{Comparative Analysis of State-of-the-Art DNN Deployment and Optimization Approaches for Heterogeneous Platforms.} 
  \label{tab:comparison}
  \begin{threeparttable}
  \setlength{\tabcolsep}{3.2pt}
  \begin{tabular}{@{}lcccccc@{}} 
  \toprule
  Approach & \makecell{DVFS- \\ {Aware}} & \makecell{Tiling- \\ Aware } & \makecell{Deadline- \\ Aware} & \makecell{Energy- \\ Aware} & \makecell{DNN-\\ Agnostic} & \makecell{Kernel-level  \\ Granul.} \\
  \midrule
  MATCH \cite{hamdi2024match}       & No              & Yes            & No        & Yes(Trd.)\tnote{i} & No & Yes \\
  RoaD-RuNNer \cite{kakolyris2023road} & No              & No             & No        & Yes(Trd.)& No & No \\
  AdaKnife \cite{liu2024adaknife}    & Yes(App)        & No\tnote{a}    & No        & No         & Yes & No\tnote{b} \\
  Kang et al. \cite{kang2020scheduling}   & Yes(App)        & No             & Yes       & Yes(Opt.)\tnote{i}&  No & No \\
  DORY \cite{burrello2021dory}        & No              & Yes            & No        & No         & No & N/A\tnote{f} \\
  {FDT} \cite{stahl2023fused} & No              & Yes            & No        & No         & No & N/A\tnote{f} \\
  ODIMO \cite{risso2025optimizing}     & No              & N/S\tnote{d}   & No        & Yes(Trd.)& No & Yes \\
  Wang et al. \cite{wang2020neural}    & No              & No             & No        & No         & No & No \\
  Vasiliadis et al. \cite{vasiliadis2022best} & No              & No             & No        & Yes(Trd.) & Yes & No \\
  AxoNN \cite{dagli2022axonn}       & No              & No             & No        & Yes(Trd.)& No & No \\
  Map-and-Conquer \cite{bouzidi2023map}& Yes(Layer)      & No             & Yes       & Yes(Opt.) & Yes & No\tnote{g} \\
  TensorRT (Jeong) \cite{jeong2021deep} & No              & No             & No        & No         & No & No \\
  DART \cite{xiang2019pipelined}      & No              & No             & Yes       & No         & No & No \\
  HaX-CoNN \cite{dagli2024shared}   & No              & No\tnote{c}    & Yes       & Yes(Trd.) & No & No \\
  OmniBoost \cite{karatzas2023omniboost}& No              & No             & No        & No         & No & No \\
  MaGNAS \cite{odema2023magnas}     & Yes(App)        & No             & Yes       & Yes(Opt.)& No & No\tnote{h} \\
  \midrule
  \textbf{{\schedname} (Ours)} & \textbf{Yes(Kernel)} & \textbf{Yes(Adap.)} & \textbf{Yes} & \textbf{Yes (Opt.)} & \textbf{Yes} & \textbf{Yes}\tnote{e} \\
  \bottomrule
  \end{tabular}%
  \begin{tablenotes}[para,flushleft]
    \scriptsize
    \item[a] Cannot tile kernels that individually exceed PE memory, although network partitioning respects PE limits.
    \item[b] Mixed granularity: Kernel/layer-level for CNNs; coarser (e.g., encoder block) for Transformers.
    \item[c] Manages shared memory contention/throughput, but not PE memory capacity limits.
    \item[d] Memory management details not specified, despite hardware-aware PE selection.
    \item[e] Supports any DNN composed of supported kernels (e.g., Conv, MatMul, FFT, Add, Norm); evaluated on Transformers.
    \item[f] Focuses only on tiling generation for specified DNN types, not full runtime execution management.
    \item[g] Partitions layers width-wise (e.g., CNN channels, Transformer heads) for mapping to CUs.
    \item[h] Scheduling occurs at the GNN block level (e.g., Grapher, FFN), specific to GNNs.
    \item[i] Trd.: Considers energy-speedup tradeoff. Opt.: Optimizes energy for a given deadline.
  \end{tablenotes}
  \end{threeparttable}
\end{table*}

The efficient management of DNNs on heterogeneous edge platforms is an active and multifaceted research area. 
This review analyzes a wide range of state-of-the-art approaches to resource management to gain a comprehensive overview. It extends beyond end-to-end schedulers to include specialized optimization frameworks. 
We consider works focused primarily on ULP's memory optimization through automated tiling ({DORY} \cite{burrello2021dory}, {FDT} \cite{stahl2023fused}), and design-time compilation and training-time mapping for HULPs ({MATCH} \cite{hamdi2024match}, {ODIMO} \cite{risso2025optimizing}). 
Targetting edge, yet still resource-aware, platforms, we also examine runtime schedulers and partitioners targeting diverse objectives such as energy-latency tradeoffs ({Kang et al.} \cite{kang2020scheduling}, {AxoNN} \cite{dagli2022axonn}, {Map-and-Conquer} \cite{bouzidi2023map}), 
real-time guarantees ({DART} \cite{xiang2019pipelined}), throughput maximization ({TensorRT} \cite{jeong2021deep}, {OmniBoost} \cite{karatzas2023omniboost}, \cite{wang2020neural}), multi-DNN contention ({HaX-CoNN} \cite{dagli2024shared}), and architecture-mapping co-design ({MaGNAS} \cite{odema2023magnas}).
Despite valuable contributions from these varied domains, our analysis reveals significant gaps in providing a holistic solution for HULPs, 
particularly concerning the integration of fine-grained, DVFS-, tiling-, and deadline-aware energy-optimized manager. 
Table~\ref{tab:comparison} provides a comparative summary, and the following paragraphs will analyze these key features in detail.

\paragraph{Scheduling Granularity and DNN Architecture Support}
The effectiveness of a DNN manager on HULPs is largely determined by its scheduling granularity, the fundamental unit of work it manages. 
We define scheduling granularity by the size of the computational unit managed: a \textit{layer} (e.g., a complete convolutional layer or Transformer block) is a high-level structural component, whereas a \textit{kernel} (or operator, e.g., matrix multiplication, 2D convolution, an activation function) is a fundamental mathematical operation within such layers. 
We argue that a fine, kernel-level granularity is paramount on HULPs, as it not only allows for precise resource control but also provides inherent support for diverse DNN architectures, a feature lacking in many specialized, architecture-centric solutions. 
However, the one limitation of many existing approaches is their \emph{architecture-centric design}. 
The majority of these frameworks are \textit{CNN-centric} \cite{hamdi2024match, kakolyris2023road, kang2020scheduling, wang2020neural, risso2025optimizing, dagli2022axonn, jeong2021deep, xiang2019pipelined, dagli2024shared, karatzas2023omniboost}, effectively making them specialized CNN managers rather than general DNN solutions.
This specific focus is also apparent in other architecture-centric managers like MaGNAS \cite{odema2023magnas} for GNNs. 
The other limitation is the \textit{coarse scheduling granularity} of many existing approaches, often align with high-level structural components in comparison to individual kernels, such as entire inferences \cite{wang2020neural, vasiliadis2022best} or a group of computations at different levels \cite{xiang2019pipelined, dagli2024shared, kang2020scheduling, dagli2022axonn, jeong2021deep, karatzas2023omniboost, liu2024adaknife, risso2025optimizing, bouzidi2023map},
for example, grouping such as width-wise splits of attention heads or channels \cite{risso2025optimizing, bouzidi2023map}. 
Thus, a significant gap remains for solutions that leverage deep kernel-level control to provide inherent unified support for diverse DNN architectures on HULP platforms.

\paragraph{{DVFS and its Granularity}}
A key differentiator in energy-efficient resource management and scheduling is the utilization of {DVFS}, a fundamental power management technique that achieves significant energy savings by dynamically adjusting PE voltage and frequency to match the target workload, 
leveraging the quadratic power-voltage relationship ($P \propto V^2 f$) \cite{liu2021dynamic, zidar2024dynamic}.
While hardware platforms like Zero-riscy \cite{nunez2023risco2} and Raven \cite{keller2017risc} demonstrate its feasibility {of runtime DVFS control in RISC-V based ULPs}, 
our review reveals that {existing approaches utilize this capability at varying levels of granularity, if at all.}
{A significant majority of the analyzed related works do not incorporate any form of DVFS.} 
This includes approaches focused on compilation (MATCH \cite{hamdi2024match}, DORY \cite{burrello2021dory}, FDT \cite{stahl2023fused}), 
partitioning/offloading (RoaD-RuNNer \cite{kakolyris2023road}), training-time mapping (ODIMO \cite{risso2025optimizing}), 
runtime scheduling primarily for throughput or contention management (TensorRT-based \cite{jeong2021deep}, DART \cite{xiang2019pipelined}, HaX-CoNN \cite{dagli2024shared}, 
OmniBoost \cite{karatzas2023omniboost}), adaptive device selection \cite{vasiliadis2022best}, and energy-aware layer mapping \cite{dagli2022axonn}. Even works characterizing platform performance often neglect to detail DVFS utilization \cite{wang2020neural}. 
These approaches cannot dynamically adjust operating points to minimize energy based on the immediate needs of different DNN kernels, often defaulting to fixed performance settings.
Several works consider DVFS, but typically at a coarse granularity throughout the application. 
For example, MaGNAS \cite{odema2023magnas} selects a fixed V-F during its design-time search; Kang et al. \cite{kang2020scheduling} account for the effects of V-F in its profiling; 
and AdaKnife \cite{liu2024adaknife} explores static frequency adjustments. 
Map-and-Conquer \cite{bouzidi2023map} incorporates DVFS as an optimization parameter in its coarse-layer granularity scheduling.
\begin{cmodb}
Related compiler work by Hayashi et al.~\cite{hayashi2011oscar} extends the OSCAR parallelizing compiler and API to heterogeneous multicores.
Their framework starts from sequential C/Fortran programs with accelerator hints, decomposes the program into coarse-grain macro-tasks (basic blocks, loops, and function/subroutine calls), statically maps these macro-tasks to CPUs or accelerators while considering data-transfer overheads, and applies compiler-controlled DVFS and clock gating to satisfy real-time deadlines on the RP-X heterogeneous SoC.
This is an important precedent for combining heterogeneous task scheduling with frequency selection.
However, it targets automatic parallelization and code/API generation for general programs on a higher-performance multimedia SoC, whereas {\schedname} targets design-time energy minimization for DNN inference on HULP platforms.
In particular, {\schedname} assumes a profiled DNN kernel list, schedules at kernel/operator granularity, jointly selects PE assignment, per-kernel V--F point, and memory-aware adaptive tiling, and solves the resulting deadline-constrained MCKP/ILP under tight PE-local memory constraints.
\end{cmodb}
This general absence of dynamic, fine-grained, kernel-level DVFS in the literature indicates a significant missed opportunity for energy savings on HULPs.

\paragraph{{Tiling-Awareness for Constrained Memory}}
A few approaches, despite lacking other features, such as kernel-level DVFS scheduling or strict deadline-driven optimization, provide effective solutions specifically for tiling and memory management. 
For instance, {DORY} \cite{burrello2021dory} uses constraint programming to optimize 
tiling for limited SRAM on PULP devices. {FDT} \cite{stahl2023fused} introduces a specialized depthwise tiling method specifically for TinyML memory optimization. 
The MATCH \cite{hamdi2024match} compilation framework employs Design Space Exploration (DSE) to determine effective tiling strategies based on HW models.
However, many other approaches lack comprehensive memory awareness of individual PE capacity limits. Although some consider related aspects such as memory contention
between concurrent DNNs (HaX-CoNN \cite{dagli2024shared}), data transfer overheads (\cite{kang2020scheduling, wang2020neural, vasiliadis2022best, dagli2022axonn, bouzidi2023map, odema2023magnas}), 
or aim for reduced memory usage (AdaKnife \cite{liu2024adaknife}), they do not implement tiling specifically to fit operations exceeding PE memory capacity. 
Several other frameworks do not appear to incorporate PE memory constraints or tiling mechanisms at all (e.g., \cite{kakolyris2023road, jeong2021deep, xiang2019pipelined, karatzas2023omniboost}). 
With the exception of works mainly focused on tiling strategies or memory-centric compilation, many existing approaches do not deeply integrate robust, memory-aware tiling based on PE capacity
into their core decision-making processes, representing a critical gap for practical deployment.

\paragraph{Deadline-Awareness and Energy Optimization}
Effective resource management and scheduling for HULPs require simultaneously considering application time constraints (deadlines) and optimizing energy efficiency,
often involving an explicit energy performance trade-off.
Several works incorporate these aspects to varying degrees, though often with limitations in other critical areas like granularity or dynamic power management.
Some approaches explicitly target both: Kang et al. \cite{kang2020scheduling} employ a GA-based scheduler that considers deadlines (via penalty) and allows users to select energy-latency trade-off points. 
{Map-and-Conquer} \cite{bouzidi2023map} includes target latency constraints and energy modeling. 
{MaGNAS} \cite{odema2023magnas} co-optimizes GNN architecture and mapping under latency constraints while considering energy tradeoffs. 
{HaX-CoNN} \cite{dagli2024shared} implicitly addresses timing via QoS requirements while considering PE power consumption. 
Other works prioritize meeting deadlines without deep energy optimization, such as {DART} \cite{xiang2019pipelined} for real-time tasks. 
In contrast, several frameworks explore energy vs. performance trade-offs or specific energy targets but without strictly adhering to application deadlines. 
For instance, {AxoNN} \cite{dagli2022axonn} minimizes time under an energy budget, {RoaD-RuNNer} \cite{kakolyris2023road} finds Pareto optimal energy/latency points, 
and {ODIMO} \cite{risso2025optimizing} balances accuracy and efficiency. 
Frameworks like {Vasiliadis et al.} \cite{vasiliadis2022best}, {MATCH} \cite{hamdi2024match}, and the {TensorRT-based} approach \cite{jeong2021deep} 
offer options to minimize energy or latency but lack a mechanism to enforce a strict timing constraint during optimization.
Frameworks centered on memory optimization through tiling (e.g., {DORY} \cite{burrello2021dory}, {FDT} \cite{stahl2023fused}), 
throughput maximization (e.g., {OmniBoost} \cite{karatzas2023omniboost}), or partitioning/offloading latency (e.g., {AdaKnife} \cite{liu2024adaknife}) 
often lack explicit mechanisms for managing energy consumption within a specific timing budget. 
Even works characterizing platform performance \cite{wang2020neural} typically do not go into deadline-driven scheduling. 
%

\paragraph{Summary of Gaps and Proposed Approach}
In summary, our review (Table~\ref{tab:comparison}) highlights several significant gaps in current resource management and scheduling approaches when applied to HULPs. 
These include a prevalent lack of kernel-level DVFS, insufficient memory-aware tiling, limited focus on optimizing energy under strict timing constraints, 
{and} coarse scheduling granularities unsuitable for specialized accelerators and restricted DNN architecture support. 
Collectively, these limitations hinder the ability to fully utilize the potential of HULP platforms for energy-efficient execution of DNN-based applications.
Although individual works may address some of these aspects, a holistic solution that effectively integrates all these demanding requirements for the ULP domain has been notably absent. 
To address this confluence of challenges, 
this paper proposes {{\schedname}}, a novel design-time multi-objective manager that integrates kernel-level scheduling, kernel-level DVFS, 
memory-aware adaptive tiling within a flexible, timing-constrained energy optimization strategy explicitly tailored for HULPs.

%% file: secs/sched.tex
\begin{figure}[htbp]
    \centering
    \includegraphics[width=1\linewidth]{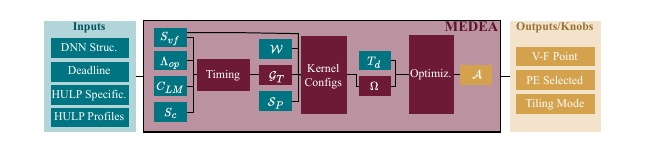}
    \caption{Overview of the {\schedname} design-time multi-objective manager, illustrating its inputs, output/knobs, and core management and optimization.}
    \Description{Detailed MEDEA workflow showing workload, deadline, and platform-profile inputs; timing, power, and tiling models; MCKP optimization; and the resulting per-kernel schedule.}
    \label{fig:sched}
\end{figure}

This section details {\schedname}, 
our proposed design-time multi-objective manager designed for energy-efficient, latency-constrained DNN inference on {HULPs}. 
Figure~\ref{fig:sched} provides a high-level overview of the {\schedname} manager, illustrating its key inputs (application representation, platform characteristics, timing constraints), 
the control knobs it manipulates (PE assignment, V-F settings, tiling modes), {and} its core management and optimization algorithm. Each of these aspects will be described in detail in the following subsections.

\subsection{{\schedname} Input Requirements} \label{sec:manger:inputs}
{As illustrated in Figure~\ref{fig:sched}, {\schedname} relies on a set of inputs: (1) the DNN-based application structure to be scheduled, (2) its associated timing constraint, (3) HULP's specifications, 
and (4) HULP's performance profiles derived from characterization}. The following subsections detail these required inputs.

\subsubsection{Application {Structure} and Timing Constraint} \label{sec:manger:inputs:app}
{{\schedname} requires the workload structure and its timing constraint.} 
First, the target {DNN-based application structure} is presented as a sequential workload $\mathcal{W}$, which is an ordered list of $N_{\mathcal{W}}$ computational kernels:
\begin{equation}
    \mathcal{W} = \{k_1, k_2, \dots, k_{N_{\mathcal{W}}} \}
    \label{eq:workload_def}
\end{equation}
Each kernel $k_i$ within this workload is defined by a tuple $(\tau_i, s_i, \delta_i)$, where $\tau_i \in \mathcal{T}_{ops}$ denotes the type of kernel, e.g., $\mathcal{T}_{ops} = \{\texttt{matmul}, \texttt{conv2d}, \texttt{norm}, \texttt{add}, \dots\}$; $s_i$ represents its operational size or dimensions (e.g., matrix dimensions for \texttt{matmul}); and $\delta_i$ specifies the datawidth (e.g., int8, int16). 
This kernel-level representation {is fundamental to} the fine-grained optimization performed by {\schedname}.
Helper utilities are provided to help generate $\mathcal{W}$ from higher-level descriptions (e.g. DNN model layers).
\cmod{The list order of $\mathcal{W}$ encodes a valid \emph{topological execution order} of the underlying DNN computation graph: kernel $k_i$ appears only after all operators whose outputs it consumes.
{\schedname} does not reorder kernels; it assumes sequential execution---$k_i$ completes before $k_{i+1}$ begins---which matches the orchestration model on severely memory-constrained HULPs (Sec.~\ref{sec:setup:platform}).}
Coarser-grained scheduling approaches, considered for comparison, typically group consecutive kernels from $\mathcal{W}$ into larger blocks. 
{The second input is the overall application timing constraint, or deadline $T_d$. 
This deadline dictates the maximum allowable active execution time for the entire workload $\mathcal{W}$ and serves as the {trade-off point} for the energy optimization performed across the sequence of kernels.}
\cmod{For the HULP deployment class targeted in this work, the intended application input is one workload~$\mathcal{W}$ with one deadline~$T_d$: {\schedname} optimizes a single, latency-constrained DNN inference at design time (Sec.~\ref{sec:limitations:scope}).}


\subsubsection{{HULP's Specifications}} \label{sec:manger:inputs:plat_static} 
%
This begins with defining the set $\mathcal{P}$ of $N_{\mathcal{P}}$ distinct PEs:
\begin{equation}
    \mathcal{P} = \{p_1, p_2, \dots, p_{N_{\mathcal{P}}} \}.
    \label{eq:platform_def}
\end{equation}
Thus, $\mathcal{P}$ typically includes one or more general-purpose CPUs (e.g., $p_{cpu}$) and various specialized accelerators (e.g., $p_{acc1}$). 
The platform operates across a discrete set $\mathcal{V}$ of $N_{\mathcal{V}}$ voltage levels, with available V-F operating points forming a set $S_{vf}$:
\begin{equation}
    S_{vf} = \{ (v_l, f_l) \mid v_l \in \mathcal{V}, l=1, \dots, N_{\mathcal{V}} \},
    \label{eq:voltage_freq_def}
\end{equation}
where, consistent with common practice in works like \cite{taji2024energy}, the system {is usually operating} at the maximum supported frequency, $f_l = F_{max}(v_l)$, at voltage $v_l$.

The platform's memory hierarchy is also important to know where data is available to be processed and how moving data between different levels should be handled.
Beyond off-chip flash for global DNN weights, this {can} include a shared memory resource (e.g., L2 cache or main on-chip memory) with capacity $C_{M}$, and/or private local memories (LMs) for each PE $p_j \in \mathcal{P}$.
PEs are assumed to operate primarily on data residing in their respective LMs.
The set of all LM capacities is given by:
\begin{equation}
    \mathcal{C}_{LM} = \{(p_j, C_{LM_j}) \mid p_j \in \mathcal{P} \}.
    \label{eq:local_memory_def}
\end{equation}
Furthermore, {\schedname} considers kernel-PE operational constraints, where each PE $p_j$ may impose limitations on specific kernel types $\tau_i$ 
(e.g., maximum supported matrix dimensions for a \texttt{matmul} kernel). These specific constraint values are denoted as $\lambda_{p_j,\tau_i}$, forming the set $\Lambda_{op}$:
\begin{equation}
    \Lambda_{op} = \{ (p_j, \tau_i, \lambda_{p_j,\tau_i}) \mid p_j \in \mathcal{P},\tau_i \in {\mathcal{T}_{ops}} \}.
    \label{eq:kernel_pe_constraints_def}
\end{equation}
These specifications define the fixed capabilities and limitations of the hardware that {\schedname} must operate within.

\subsubsection{{HULP's Performance Profiles}}\label{sec:manger:inputs:plat_prof} 
{Additionally} {\schedname} utilizes dynamic performance profiles derived from representative kernel executions on the target platform. {These profiles provide the performance metrics that form the basis for {\schedname}'s optimization process.}
First, \textit{Timing Profiles ($S_c$)} detail measured cycle counts for representive kernels $k_i$ (based on its operational size $s_i$ and data width $\delta_i$) on each relevant PE $p_j$. 
These profiles contain processing-only cycles and data movement characteristics that {\schedname} utilizes alongside $\Lambda_{op}$ and $C_{LM_j}$ to estimate cycle counts for the different tiled execution modes (Sec.~\ref{sec:manger:outputs}).
Second, \textit{Power Profiles ($S_P$)} supply characterized power information for each kernel $k_i$ executing on PE $p_j$ across different voltage levels $v_l$. This includes separated static ($P_{stat}$) and dynamic ($P_{dyn\_base}$ at a reference frequency {$f_{base}$}) power components. 
A common technique to decouple these components involves {measuring at least in two different frequencies} for a given voltage \cite{machetti2024x}. Lastly, the global idle / sleep power of the platform, $P_{slp}$, must also be profiled and provided. 

\subsection{{\schedname} Outputs/Knobs} \label{sec:manger:outputs}
    Figure~\ref{fig:sched} illustrates the outputs or control knobs that {\schedname} manipulates to generate an energy-efficient, per-kernel decisions ($\mathcal{A}$) for any kernel in the given DNN workload ($\mathcal{W}$) on the target HULP. 
    Firstly, {\schedname} performs \textit{PE Assignment}, selecting the specific PE $p_j \in \mathcal{P}$ for the execution of each kernel $k_i$. 
    This decision leverages the platform's heterogeneity by matching kernel characteristics to PE strengths. 
    Secondly, for each kernel $k_i$ on the assigned PE, {\schedname} determines the {efficient} \textit{V-F Setting} $(v_l, f_l)$ from the set $S_{vf}$, enabling kernel-level DVFS.

    Thirdly, {\schedname} chooses the \textit{Tiling Mode}, critical for managing limited local PE memory ($C_{LM_j}$) and adhering to kernel-PE operational constraints ($\lambda_{p_j,\tau_i}$ from $\Lambda_{op}$).
    Typically, for the execution of $k_i$ on a PE $p_j$, operands are transferred from a shared memory tier (e.g., L2 cache) to $p_j$'s Local Memory (LM).
    This movement often necessitates \textit{tiling} when a kernel's data requirements exceed either the LM capacity or other kernel-PE operational limits.
    Tiling decomposes $k_i$ into smaller, manageable chunks (tiles) whose data footprint adheres to both $C_{LM_j}$ and specific constraints $\lambda_{p_j,\tau_i}$.

    {when implementing tiling, {\schedname}'s role is to decide how to overlap data movement with computation efficiently. It considers two tiling modes, \textit{Single-Buffer Tiling ($t_{sb}$)} and \textit{Double-Buffer Tiling ($t_{db}$)}. 
    $t_{db}$ aims to hide data movement latency by overlapping the computation of the current tile with the data transfer of the next/previous tile, typically using roughly half of $C_{{LM}_j}$ for computation buffers. 
    In contrast, $t_{sb}$ maximizes the tile size based on constraints {($C_{LM_j}, \lambda_{p_j,\tau_i}$)}. Consequently, it has zero overlap between data movement and computation, processing tiles sequentially.} 
    {\schedname} selects the most efficient mode for each kernel on its assigned PE.

\subsection{{\schedname} Management and Optimization Algorithm} \label{sec:manger:algo}
The primary objective of {\schedname} is to generate a schedule that minimizes the total energy consumption, $E_t$, for executing a given DNN-based application workload $\mathcal{W}$ (defined in Eq.~\eqref{eq:workload_def}) within a specified timing constraint, deadline $T_d$. 
This total energy, $E_t$, comprises the energy consumed during active DNN execution, $E_{t,a}$, and the energy consumed during any subsequent idle/sleep period, $E_{t,s}$, until the next operational cycle.
\begin{equation}
    E_t = E_{t,a} + E_{t,s}.
    \label{eq:total_energy_components} 
\end{equation}
%
Taking into account idle power $P_{slp}$ and active execution time $T_{t,a} (\le T_d)$, the total energy to be minimized is:
\begin{equation}
    E_{t} = E_{t,a} + P_{slp} \times \max(0, T_d - T_{t,a}).
    \label{eq:total_energy_calculation}
\end{equation}

\paragraph{Timing and Power}
{To make its decisions, {\schedname} leverages characterized timing ($S_c$) and power ($S_P$) information. 
A timing model, $\mathcal{G}_{T}(\cdot)$, is employed to estimate execution times.
This model utilizes measured cycle counts ($S_c$) for computation, which are obtained for each kernel $k_i$ on a specific PE $p_j$. 
The model first estimates the total cycle count for a given kernel and a chosen tiling mode $t_m \in \{t_{sb}, t_{db}\}$. 
It includes both directly profiled processing-only cycles, extrapolated values for non-profiled kernel sizes, as well as data movement cycles determined by the chosen tiling mode ($t_m$) and platform constraints ($C_{LM_j}, \Lambda_{op}$).
Subsequently, this total cycle count is converted to actual execution time by dividing by the operational frequency $f_l$, which is derived from the selected voltage level $v_l$ and the platform's frequency-voltage mapping $S_{vf}$. 
Thus, $\mathcal{G}_{T}(\cdot)$ provides the estimated execution time of a kernel $k_i$ on a specific PE $p_j$ at a voltage level $v_l$ and using a tiling mode $t_m$.
For power, the model directly uses the characterized power profiles ($S_P$). It is assumed that the power consumption for a given kernel type $\tau_i$ on PE $p_j$ at voltage $v_l$ is independent of the kernel's operational size $s_i$.}

\paragraph{Kernel Execution Configurations}
The total active energy, $E_{t,a}$, is the sum of the energies consumed by individual kernels $k_i \in \mathcal{W}$. 
For each kernel $k_i$, {\schedname} controls several knobs: the PE assignment $p_{i} \in \mathcal{P}$, the voltage $v_{i}$ (which determines $f_{i}$ from $S_{vf}$), and the execution/tiling mode $c_{i} \in \{t_{sb}, t_{db}\}$ (as detailed in Section~\ref{sec:manger:outputs}). 
A specific combination of these choices for kernel $k_i$ forms an execution configuration, denoted as $\omega_{ij} = (p_{ij}, v_{ij}, c_{ij})$, where $j$ is an index over the possible valid configurations for kernel $i$.
Let $\Omega_i$ be the set of all such valid configurations for kernel $k_i$. A configuration is deemed valid if its execution time $\mathcal{G}_{T}(\omega_{ij})$ can be successfully estimated (e.g., it is not an invalid mapping due to the PE capability or constraints).
The active time and active energy for kernel $k_i$ executed with a configuration $\omega_{ij}$ are:
\begin{align}
    T_{a}(\omega_{ij}) &= \mathcal{G}_{T}(\omega_{ij}) \label{eq:kernel_time_def}, \\
    E_{a}(\omega_{ij}) &= \mathcal{G}_{P}(\omega_{ij}) \times T_{a}(\omega_{ij}). \label{eq:kernel_energy_def}
\end{align}

\paragraph{Optimization Objective Rationale}
{The overall goal is to minimize the total energy $E_t$ (Eq.~\eqref{eq:total_energy_calculation}) while ensuring the total active time $T_{t,a}$ respects the deadline $T_d$. 
This comprehensive objective can be simplified. Consider any valid schedule that meets the deadline with an active time $T_{t,a}$ and active energy $E_{t,a}$. 
Any alternative schedule that executes faster ($T'_{t,a} < T_{t,a}$) would necessarily require higher V-F settings for some kernels, 
resulting in a higher active energy consumption ($E'_{t,a} > E_{t,a}$). Furthermore, if the idle power $P_{slp}$ is greater than zero, this faster schedule would also 
incur a longer subsequent idle period ($T_d - T'_{t,a}$), leading to a higher idle energy consumption ($E'_{t,s} > E_{t,s}$). 
Since executing faster increases both the active and (for $P_{slp}>0$) idle energy components, there is no benefit to finishing earlier than necessary. 
The optimal strategy is therefore to find the schedule that utilizes the available time up to the deadline $T_d$ with the lowest possible active energy. 
Consequently, the optimization problem simplifies to minimizing the total active energy $E_{t,a}$ subject to the constraint $T_{t,a} \le T_d$.}

\paragraph{Optimization Formulation and Solution}
The optimization problem involves selecting one configuration $\omega_{ij}$ from $\Omega_i$ for each kernel $k_i$.
Before generating the full set $\Omega_i$, for each kernel $k_i$ and each potential PE and V-F pair $(p_{ij}, v_{ij})$, {\schedname}, utilizing $\mathcal{G}_{T}(\cdot)$, pre-selects the execution mode $c_{ij}$ (from $\{t_{sb}, t_{db}\}$) that yields the minimum execution cycles (and thus minimum $T_a(\omega_{ij})$ for that $(p_{ij}, v_{ij})$). 
This step effectively reduces the dimensionality of the choices before the main optimization. 
We also define binary decision variables $x_{ij}$, where $x_{ij}=1$ if the $j$-th configuration $\omega_{ij} \in \Omega_i$ is selected for kernel $k_i$, and $0$ otherwise.
\cmod{Because execution order is fixed by $\mathcal{W}$, the formulation does not introduce explicit inter-kernel precedence constraints: the ILP/MCKP selects one execution configuration per kernel, and total active time $T_{t,a}$ is the sum of per-kernel durations under the assumption of non-overlapping kernel execution.}
The optimization problem is to \textit{minimize} total active energy $E_{t,a}$:
\begin{equation}
    E_{t,a} = \sum_{i=1}^{N_{\mathcal{W}}} \sum_{\omega_{ij} \in \Omega_i} E_a(\omega_{ij}) \times x_{ij}
    \label{eq:ilp_objective_final}
\end{equation}
\textit{Subject to:}
\begin{enumerate}

    \item \textit{Timing Constraint:} Let $T_{t,a} = \sum_{i=1}^{N_{\mathcal{W}}} \sum_{\omega_{ij} \in \Omega_i} T_a(\omega_{ij}) \times x_{ij}$. Then,
    \begin{equation}
    T_{t,a} \le T_d \label{eq:constraint_timing_final}
    \end{equation}

    \item \textit{Unique Assignment:} Each kernel must be assigned exactly one configuration:
    \begin{equation}
    \sum_{\omega_{ij} \in \Omega_i} x_{ij} = 1 \quad \forall i \in \{1, \dots, N_{\mathcal{W}}\}
    \label{eq:constraint_unique_final}
    \end{equation}

    \item \textit{Binary Decision:} Decision variables are binary:
    \begin{equation}
    x_{ij} \in \{0, 1\} \quad \forall i \in \{1, \dots, N_{\mathcal{W}}\}, \forall \omega_{ij} \in \Omega_i
    \label{eq:constraint_binary_final}
    \end{equation}
\end{enumerate}
This optimization problem is then assigned and solved as a Multiple Choice Knapsack Problem (MCKP). 
In this mapping, each kernel $k_i$ corresponds to an item group, and its valid configurations $\omega_{ij} \in \Omega_i$ represent the items to choose from within that group. 
The active energy $E_a(\omega_{ij})$ (Eq.~\eqref{eq:kernel_energy_def}) is the \textit{value} or cost associated with selecting item $\omega_{ij}$ (to be minimized), 
its execution time $T_a(\omega_{ij})$ (Eq.~\eqref{eq:kernel_time_def}) is its \textit{weight}, and the application deadline $T_d$ serves as the knapsack's \textit{capacity}. 
{\schedname} solves this MCKP formulation using an Integer Linear Programming (ILP) solver via the \texttt{PuLP} library in Python \cmod{with its default CBC (\textit{Coin-or Branch and Cut}) backend}.
\cmod{This solver runs once on a host during deployment preparation and is never invoked on the HULP during inference; on-device optimization overhead is therefore zero.}

\paragraph{Schedule Extraction}
The ILP solver yields the optimal set of $x_{ij}$ values. 
{\schedname} then extracts the chosen configuration $\omega^*_{i}$ (which encapsulates the selected $p^*_{i}, v^*_{i},$ and pre-optimized $c^*_{i}$) for each kernel $k_i$ for which $x_{ij}=1$. 
This sequence of configurations, $\mathcal{A} = \{\omega^*_{1}, \omega^*_{2}, \dots, \omega^*_{N_{\mathcal{W}}}\}$, constitutes the final, kernel-level, energy-efficient schedule generated by the manager.
\cmod{At runtime, the host CPU executes~$\mathcal{A}$ strictly in list order---kernel~$k_i$ completes before~$k_{i+1}$ begins---matching the sequential orchestration model on HEEPTimize (Sec.~\ref{sec:setup:platform}).}
The overall {\schedname} scheduling process is depicted in Figure~\ref{fig:sched}.


%% file: secs/setup.tex
 This section details the experimental methodology for evaluating {\schedname}. 
 In particular, it covers the evaluation platform including its architecture, implementation and characterization; the software toolchain; the application case study with its experimental conditions; and the baseline approaches used for comparison.
 
 \subsection{Evaluation Platform: HEEPTimize} \label{sec:setup:platform}
\begin{figure}[htbp]
     \centering
     \includegraphics[width=1\linewidth]{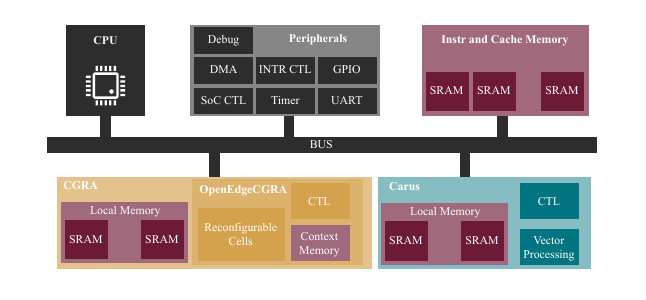}
     \caption{High-level block diagram of HEEPTimize.}
     \Description{HEEPTimize architecture with a RISC-V host, shared memory and interconnect, and attached Carus near-memory-computing and OpenEdgeCGRA processing elements.}
     \label{fig:heeptimize_overview}
 \end{figure}

 %
 %
 {To evaluate {\schedname} in a HULP environment, we use the X-HEEP platform \cite{x-heep_1}, which is a configurable and extensible open source platform
 designed specifically for accelerator prototyping for edgeAI applications. 
 We use its eXtendible Accelerator InterFace (XAIF) to integrate the accelerators tested with {\schedname}. 
 We name the resulting prototyping platform, which has been synthesized for prototyping on a Xilinx Zynq UScale+ FPGA and 
 for power characterization using an ASIC flow, \emph{HEEPTimize} (Figure~\ref{fig:heeptimize_overview}).}
 
 \subsubsection{Architecture} \label{sec:heeptimize:arch}
 
 HEEPTimize integrates a low-power RISC-V processor with two types of specialized accelerators: an CGRA and an NMC. 
 {These three PEs correspond to the set $\mathcal{P}$, Eq.~\eqref{eq:platform_def}, considered by {\schedname}.}
 This selection stems from their demonstrated potential for high energy efficiency and computational flexibility in ULP applications \cite{caon2024scalable, alvarez2023open}. 
 
 %
 The main controller is the {CV32E40P} \cite{schiavone2017slow}, an open-source, 32-bit, 4-stage, in-order RISC-V core (RV32IMC ISA) optimized for low power. 
 {It handles general-purpose tasks, orchestrates accelerator operations, and executes kernels that are not efficiently
  supported by the specialized units. 
  While OpenEdgeCGRA and Carus excel at some common arithmetically intensive DNN kernels (e.g., \texttt{matmul}, \texttt{conv2d}, \texttt{add}, \texttt{norm}), 
  they are less optimized for complex nonlinear functions or floating-point arithmetic. Consequently, operations such as Softmax, the standard GeLU activation, and floating-point calculations 
  like the logarithm of an FFT's amplitude are typically offloaded to the host CPU, motivating the hardware-aware model optimizations detailed in Section~\ref{sec:setup:app}.}
 The first accelerator is {OpenEdgeCGRA} \cite{alvarez2023open}, a flexible CGRA. 
 Its architecture comprises a four-by-four mesh of identical reconfigurable cells (RCs) in a torus topology, each equipped with a 32-bit ALU, registers, and private program memory. 
 These RCs operate in a time-multiplexed fashion based on column program counters. Via XAIF, OpenEdgeCGRA utilizes four 32-bit master ports for data memory access and two 32-bit slave ports for its configuration and instruction memory.
 The second accelerator is {Carus} \cite{caon2024scalable}, an NMC unit optimized for vector operations. It integrates a programmable RISC-V eCPU controller with a scalable Vector Processing Unit (VPU), 
 executing custom \texttt{xvnmc} ISA vector operations directly on its Vector Register File (VRF), configured with four SRAM banks. 
 The eCPU manages VPU operations using kernel code loaded by the host into a dedicated eMEM, and Carus natively supports 8-bit, 16-bit, and 32-bit fixed-point data. Reflecting its NMC design, 
 Carus connects via a single XAIF slave port for eMEM programming and control, which lacks external master ports. 
 The completion of the computation for both OpenEdgeCGRA and Carus is signaled to the host CPU via XAIF-routed interrupt lines to the {interrupt controller}. 
 
 The memory hierarchy is designed to support these PEs. 
 Carus operates on its internal \SI{64}{\kibi\byte} {LM (or VRF)}. For operational similarity and fair comparison, OpenEdgeCGRA is provided with a dedicated \SI{64}{\kibi\byte} LM. 
 These LM capacities correspond to $C_{LM_j}$ values in $\mathcal{C}_{LM}$ (Eq.~\eqref{eq:local_memory_def}). A shared \SI{128}{\kibi\byte} L2 cache acts as an intermediate data buffer between these accelerator LMs and the off-chip NAND flash memory, where {application data such as model parameters} are stored. 
 DMA controllers manage data transfers between flash and the L2 cache, and subsequently between the L2 cache and the respective accelerator LMs. This L2 cache also facilitates double buffering techniques during processing.
 
 
 \subsubsection{Implementation and Characterization} \label{sec:heeptimize:impl}
 Characterization of HEEPTimize, in the absence of fabricated silicon, was based on a combination of FPGA prototyping for timing validation and an ASIC implementation flow for power and area analysis. 
 While FPGA prototyping yields accurate cycle counts for kernel execution, precise power estimation requires simulation using industry-standard Electronic Design Automation (EDA) tools.

 \paragraph{FPGA Prototyping}
 To facilitate functional verification and obtain realistic kernel execution cycle counts, 
 the HEEPTimize architecture was implemented and prototyped on a Xilinx ZCU104 Ultrascale+ FPGA development board. 
 The cycle counts measured from the execution of representative DNN kernels on each PE on the FPGA serve as input for the timing profiles used by {\schedname} (as detailed in Section~\ref{sec:manger:inputs:plat_prof}).

 \paragraph{ASIC Implementation (GF \SI{22}{\nano\meter} FDX)}
 For accurate power and area analysis, HEEPTimize was designed using a standard cell ASIC flow targeting the GlobalFoundries (GF) \SI{22}{\nano\meter} Fully Depleted Silicon-on-Insulator (FDX) technology node, chosen for its suitability for low-power applications.
 Logic synthesis for the entire design was performed using Synopsys Design Compiler (DC), targeting the worst-case Slow-Slow Global (SSG) process corner to ensure robust timing closure across process variations. 
 On-chip memories (L2 cache, instruction memory, and local memories) were generated using the GF \SI{22}{\nano\meter} FDX S1DU SRAM compiler, selected for its ULP characteristics.

 \paragraph{Timing Analysis (Max Frequency)} 
 Static Timing Analysis (STA), performed using Synopsys PrimePower, determined the maximum operating frequency ($f_{max}$) achievable at each supported voltage level (\SI{0.5}{\volt}, \SI{0.65}{\volt}, \SI{0.8}{\volt}, and \SI{0.9}{\volt}). 
 These characterized maximum frequencies form the basis of the $S_{vf}$ set (Eq.~\eqref{eq:voltage_freq_def}) and are crucial for the DVFS model in {\schedname}. The results are presented in Table~\ref{tab:freq_voltage}.
 \begin{table}[htbp]
     \centering
     \caption{HEEPTimize Maximum Operating Frequencies vs. Voltage (GF \SI{22}{\nano\meter} FDX)}
     \label{tab:freq_voltage}
     \begin{tabular}{lcccc} 
         \toprule
         Voltage (\si{\volt})       & 0.50 & 0.65 & 0.80 & 0.90 \\
         \midrule
         Max Freq. (\si{\mega\hertz}) & 122  & 347  & 578  & 690  \\
         \bottomrule
     \end{tabular}
 \end{table}

 \paragraph{{Power Analysis}} 
 System-level post-synthesis simulations were performed using Siemens QuestaSim on the synthesized netlist to capture realistic operational behavior during the execution of representative DNN kernels. 
 Detailed switching activity information from these simulations was logged into Value Change Dump (VCD) files. 
 The synthesized netlist and the generated VCD files were then input into Synopsys PrimePower for dynamic power analysis, conducted under typical-typical (TT) operating conditions. 
 This tool calculates power consumption based on the standard cell library's power models and the captured switching activity, providing a breakdown into Static (Leakage) Power ($P_{stat}$) and Dynamic Power ($P_{dyn\_b}$ at the simulation frequency, $f_b$). 
 To obtain power profiles across different voltages ($v_l \in \mathcal{V}$), this analysis was repeated using standard cell libraries characterized for each supported voltage level (\SI{0.5}{\volt}, \SI{0.65}{\volt}, \SI{0.8}{\volt}, and \SI{0.9}{\volt}).

 \paragraph{Area Results} 
 The breakdown of the post-synthesis area for HEEPTimize, which details the contribution of various parts of the memory, processor and accelerators, and peripherals, was obtained after synthesis with Synopsys DC (SSG corner), 
 is reported in Table~\ref{tab:area_breakdown}.

 \begin{table}[htbp]
     \centering
     \caption{Post-Synthesis Area Breakdown of HEEPTimize (GF \SI{22}{\nano\meter} FDX, SSG Corner)}
     \label{tab:area_breakdown}
     \begin{tabular}{lr}
         \toprule
         Component                               & Area (\si{\milli\meter\squared}) \\
         \midrule
         CPU Subsystem                           & 0.021      \\
         Carus (NMC, incl. \SI{64}{\kibi\byte} VRF) & 0.110      \\
         OpenEdgeCGRA (Logic)                    & 0.085      \\
         CGRA Local Memory (\SI{64}{\kibi\byte})   & 0.091      \\
         L2 Cache (\SI{128}{\kibi\byte})           & 0.181      \\
         Instruction Memory (\SI{64}{\kibi\byte})  & 0.091      \\
         Peripherals                             & 0.053      \\
         \midrule
         {Total Area}                & {$\approx$0.632} \\
         \bottomrule
     \end{tabular}
 \end{table}
 
 \subsection{Software Toolchain and Compilation} \label{sec:setup:toolchain}
 All software components, including DNN kernels executed on the CPU and accelerator control code, were compiled using the 
 \textit{GNU RISC-V GCC compiler, version 11.1.0} applying the \texttt{-O2} optimization level. 
 Kernels targeted for the main RISC-V CPU (CV32E40PX) were compiled for the standard RV32IMC instruction set. 
 Kernels designated for the NMC accelerator (Carus) were developed in C and RISC-V assembly, leveraging the custom \texttt{xvnmc} vector ISA extension. 
 These were compiled using an extended version of the GNU RISC-V GCC toolchain incorporating assembler support for \texttt{xvnmc} and loaded into Carus's local instruction memory (eMEM) by the host CPU at runtime. 
 
 \subsection{Application Case Study and Experimental Condition} \label{sec:setup:app}
 
\begin{figure}[htbp]
     \centering
     \includegraphics[width=\linewidth]{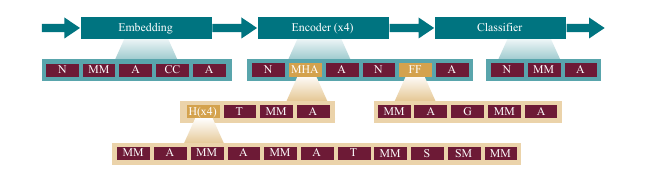}
     \caption{TSD model architecture, illustrating the decomposition into processing kernels (where G=GeLU, SM=SoftMax, T=Transpose, S=Scale, N=Norm, MM=MatMul, A=Add, CC=ClassConcat, MHA=Multi-Head Attention, FF=FeedForward, and H=Head).}
     \Description{Transformer-based seizure-detection pipeline decomposed into preprocessing, embedding, four encoder blocks with multi-head attention and feed-forward stages, and a classification head.}
     \label{fig:seizure_detection}
 \end{figure}

 \paragraph{TSD Model Description}
 We evaluated {\schedname} using the \textit{Transformer for Seizure Detection (TSD)} model \cite{amirshahi2024metawears, najafi2024versasens}. 
 This model was chosen for its relevance to ULP biomedical applications and exemplifies the trend of employing transformer architectures for complex signal 
 processing tasks on these platforms.
 The TSD model is based on a Vision Transformer (ViT) architecture specifically designed for epileptic seizure detection using EEG data. 
 It comprises four transformer encoder blocks, each containing a multi-head self-attention (MHSA) layer and a feed-forward network (FFN). 
 Figure~\ref{fig:seizure_detection} shows the detailed model architecture, illustrating its decomposition into processing kernels. 
 The model was trained and tested using the publicly available Temple University Hospital EEG Seizure Corpus (TUSZ)-v2.0.0 \cite{najafi2024versasens}.
 \cmod{We intentionally use TSD as a \emph{stress test} of the four mechanisms {\schedname} integrates on HEEPTimize: (i)~a heterogeneous kernel mix beyond homogeneous CNN stacks (Fig.~\ref{fig:seizure_detection}), including preprocessing, \texttt{matmul}/\texttt{norm}/\texttt{add} blocks, and CPU-resident nonlinearities; (ii)~memory pressure that forces nontrivial adaptive tiling given \SI{64}{\kibi\byte} accelerator local memories; (iii)~three deadline regimes ($T_d \in \{\SI{50}{\milli\second}, \SI{200}{\milli\second}, \SI{1000}{\milli\second}\}$) that expose qualitatively different scheduling trade-offs; and (iv)~a real always-on biomedical deployment mode of continuous periodic inference on wearables.}
 
 \paragraph{TSD Model Modifications}
 \cmod{%
 These changes are \textbf{upstream model--platform mapping decisions} applied before {\schedname} is invoked: they determine operator implementations for accelerator compatibility, analogously to hardware-aware adaptations in embedded transformer deployments~\cite{jung2024optimizing}.
 {\schedname} consumes the resulting ordered kernel workload~$\mathcal{W}$ and optimizes per-kernel PE assignment, voltage--frequency settings, and tiling under~$T_d$; it does \textbf{not} perform operator lowering or select among mathematically equivalent Softmax, GeLU, or FFT post-processing implementations.%
 }
 The kernels that were not compatible with the HEEPTimize's accelerators were optimized in SW using reasonable effort to facilitate that the application can be executed in an ULP.
 
 Key modifications include: replacing the original floating-point Softmax layer with a \textit{constant Softmax} approximation using a 3-coefficient Taylor expansion (following the approach in \cite{liu2024consmax}); 
 direct use of the \textit{magnitude of the fast Fourier transform (FFT)} results instead of the logarithm of the FFT amplitude used in the original work \cite{amirshahi2024metawears, najafi2024versasens}; 
 and implementation of the GeLU activation function via a \textit{piecewise linear (PWL) approximation}, replacing its standard floating-point computation.
 These modifications significantly reduce computational complexity, particularly for components not easily accelerated on HEEPTimize's specialized hardware, {making the original application more suitable for execution on ULP devices.} 
 As shown in Table~\ref{tab:modification_impact}, the CPU cycle counts for these specific operations were drastically reduced. 
 While retraining showed minimal impact on the F1-score for the Softmax and GeLU changes (\SI{66.0}{\percent} vs. \SI{66.6}{\percent}), using the FFT magnitude instead of log-amplitude resulted in a final F1-score of \SI{60.3}{\percent}, 
 deemed {an acceptable exchange for the significant reduction of \SI{98}{\percent} achieved in execution cycles on these kernels.}
 
 \begin{table}[h] 
     \centering
     \caption{CPU Cycles Reduction from TSD Model Modifications}
     \label{tab:modification_impact}
     \vspace{-2mm} 
     \begin{tabular}{lrr}
         \toprule
         Operation & Original Cycles (\si{\mega\relax}) & Modified Cycles (\si{\mega\relax}) \\
         \midrule
         Log-Amplitude FFT & $\sim$182 & $\sim$11.00 \\
         Softmax           & $\sim$647 & $\sim$5.00 \\
         GeLU              & $\sim$8   & $\sim$0.03 \\
         \bottomrule
     \end{tabular}
     \vspace{-3mm} 
 \end{table}
 
 \paragraph{Experimental Conditions}
 To comprehensively evaluate {\schedname}'s adaptability and performance, particularly for applications with stringent or varying real-time demands such as continuous inference in biomedical scenarios, 
 the TSD model (specifically its transformer core for most comparative analyses) is evaluated under three distinct timing constraints ($T_d$): \SI{50}{\milli\second} (stringent), \SI{200}{\milli\second} (moderate), and \SI{1000}{\milli\second} (relaxed). 
 These varying deadlines allow us to assess how {\schedname} manages the energy-performance tradeoff {when dealing with different application timing requirements.}
 
 \subsection{Comparison Baselines} \label{sec:setup:baselines}
 To quantify the benefits of {\schedname}, its performance is compared against several baseline scheduling {and management} strategies adapted from principles observed in the literature. 
 A necessary step for evaluation on memory-constrained platforms like HEEPTimize was the incorporation of memory management. 
 As many related works (e.g., \cite{bouzidi2023map, kakolyris2023road, odema2023magnas}) do not inherently include the required memory-aware tiling, 
 we consistently applied tiling using the double buffering ($t_{db}$) strategy (Section~\ref{sec:manger:outputs}) across all evaluated methods, ensuring feasibility and comparability on our hardware.
 \cmod{All baselines below are evaluated on the {identical modified TSD kernel list} produced by the mapping choices in Sec.~\ref{sec:setup:app}, isolating scheduling and energy-management decisions (PE selection, kernel-level DVFS, adaptive tiling) rather than operator implementation.}
 We evaluate {and compare against} the following baselines, representing increasing levels of optimization sophistication:
 \begin{itemize}
     \item {CPU (MaxVF):} A non-DVFS, homogeneous approach running the workload entirely on the CPU at its maximum V-F setting.
     \item {StaticAccel (MaxVF):} selects a priori the single most energy-efficient accelerator for the TSD workload and runs it at maximum V-F, offloading unsupported kernels to the CPU; 
     this is conceptually similar to static accelerator selection in works like \cite{wang2020neural, vasiliadis2022best}.
     \item {StaticAccel (AppDVFS):} Applies application-level DVFS to the StaticAccel (MaxVF) configuration by selecting a single optimal V-F setting for the entire execution of the chosen accelerator, similar to strategies proposed in \cite{liu2024adaknife, kang2020scheduling, odema2023magnas}.
 
     \item {CoarseGrain (AppDVFS): Represents approaches that schedule larger functional blocks by selecting the most energy-efficient PE for predefined \textit{groups} of kernels and applying a single application-level DVFS setting. 
     For the TSD model, we define these groups based on the logical components of the transformer architecture:
      The initial input embedding forms one group; within each encoder layer, the normalization, each multi-head attention head, the feed-forward network, and the residual connection are 
      each treated as separate groups; and a final group is formed for the classification block. 
      This grouping strategy reflects works that schedule larger blocks (e.g., \cite{bouzidi2023map}) while incorporating basic energy-aware PE selection (conceptually similar to \cite{hamdi2024match, risso2025optimizing}).}
 \end{itemize}

%% file: secs/results.tex
This section presents the experimental results, conducted on HEEPTimize (Section~\ref{sec:setup:platform}). We first analyze {\schedname}'s performance for running TSD model under varying timing constraints (Section~\ref{sec:setup:app}) and compare it against the defined baselines (Section~\ref{sec:setup:baselines}), and then examine its adaptive scheduling decisions.
Subsequently, we dissect the contributions of isolated {\schedname}'s key features to its overall energy efficiency.

\subsection{{\schedname} Performance Evaluation} \label{sec:results:evaluation}

\begin{figure}[htbp]
    \centering
    \includegraphics[width=\linewidth]{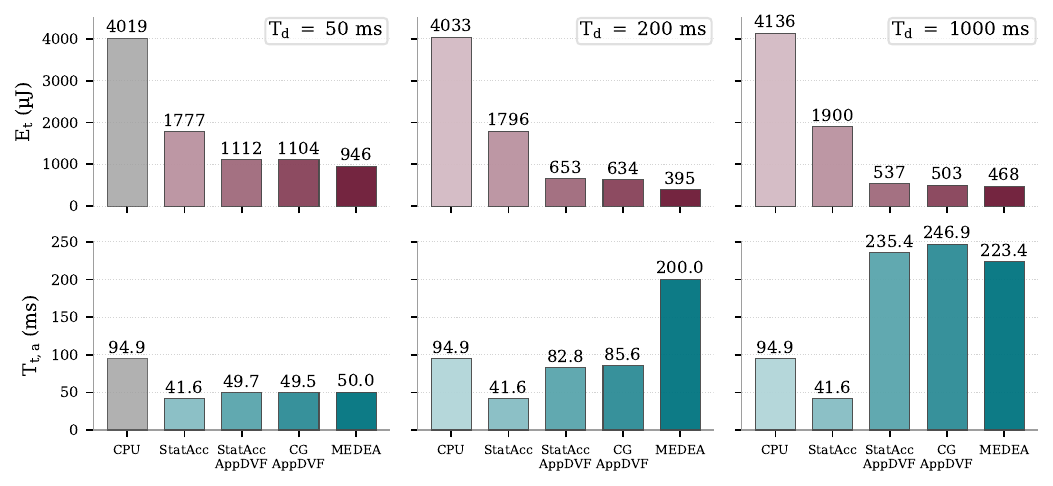} 
    \caption{Total energy and active execution time of one window while running the TSD model using {\schedname} versus baseline schedulers under timing constraints of $\SI{50}{\milli\second}$, $\SI{200}{\milli\second}$, and $\SI{1000}{\milli\second}$.}
    \Description{Grouped comparison of total energy and active execution time for MEDEA and the baseline schedulers at 50, 200, and 1000 millisecond deadlines; MEDEA meets each deadline with the lowest total energy.}
    \label{fig:overall_results}
\end{figure}

Figure~\ref{fig:overall_results} presents the total energy consumption ($E_t$) and active execution time ($T_{t,a}$) for the inference decisions generated by {\schedname} and the 
evaluated baselines (Section~\ref{sec:setup:baselines}) across the three defined timing constraints for the TSD transformer core. 
Note that $E_t$ for each scenario in Figure~\ref{fig:overall_results} represents the energy consumed over one complete inference window, with the duration of this window being defined by its corresponding deadline ($T_d$). 
Table~\ref{tab:energy_time_breakdown} further details the energy and time breakdown specifically for {\schedname}'s {decisions}, highlighting the interplay between active and idle components. 
With relaxed timing constraints (e.g., $T_d = \SI{1000}{\milli\second}$), a significant portion of the deadline window is spent idle ($T_{t,s}$), making idle energy ($E_{t,s}$) prominent. 
This underscores the importance of platform idle power ($P_{slp}$), especially when timing constraints are not demanding, as also noted in \cite{taji2024energy}.


\begin{table}[htbp]
    \centering
    \caption{End-to-End Time and Energy Breakdown for the TSD workload using {\schedname} (Sleep power $P_{slp} = \SI{129}{\micro\watt}$).}
    \label{tab:energy_time_breakdown} 
    \begin{threeparttable}
    \begin{tabular}{rrrrrr}
        \toprule
        Deadline  & Active Time  & Sleep Time  & Active Energy  & Sleep Energy  \\
        (\si{\milli\second})  & (\si{\milli\second}) & (\si{\milli\second}) & (\si{\micro\joule}) & (\si{\micro\joule}) \\
        \midrule
        50   & 50 & 0 & 946 & 0 \\
        200  & 200 & 0 & 395 & 0  \\
        1000 & 223 & 777 & 368 &  100  \\
        \bottomrule
    \end{tabular}
    \end{threeparttable}
\end{table}

As depicted in Figure~\ref{fig:overall_results}, executing the TSD model solely on the \textit{CPU} consumes excessive energy and fails to meet the $\SI{50}{\milli\second}$ deadline, confirming the need for hardware acceleration. 
The \textit{StaticAccel} baseline, while meeting deadlines, incurs high energy costs due to its lack of DVFS and inflexibility in leveraging platform heterogeneity. 
Introducing application-level DVFS via \textit{StaticAccel-AppDVF} reduces energy compared to the non-DVFS version but still underutilizes platform heterogeneity. 
Leveraging heterogeneity with \textit{CoarseGrain(CG)-AppDVF}, while an improvement over single-accelerator strategies, sees its efficiency limited by coarse granularity, a simple DVFS scheme, and the absence of explicit timing constraint-based optimization during its PE selection and AppDVFS determination.


In contrast, {\schedname} consistently optimizes energy consumption to achieve an energy-efficient solution while meeting the deadline across all evaluated scenarios. 
For instance, when compared to the \textit{CoarseGrain (AppDVFS)} baseline, {\schedname} achieves energy savings of approximately $\SI{14}{\percent}$ for the $\SI{50}{\milli\second}$ deadline, 
$\SI{38}{\percent}$ for the $\SI{200}{\milli\second}$ deadline, and $\SI{7}{\percent}$ for the $\SI{1000}{\milli\second}$ deadline, always outperforming the baselines while successfully meeting all timing constraints. 
%
This superior performance is a direct result of the integrated features of {\schedname}.

\subsection{{\schedname} Optimized Decisions} \label{sec:results:decisions}

\begin{figure}[htbp]
    \centering
    \includegraphics[width=\linewidth]{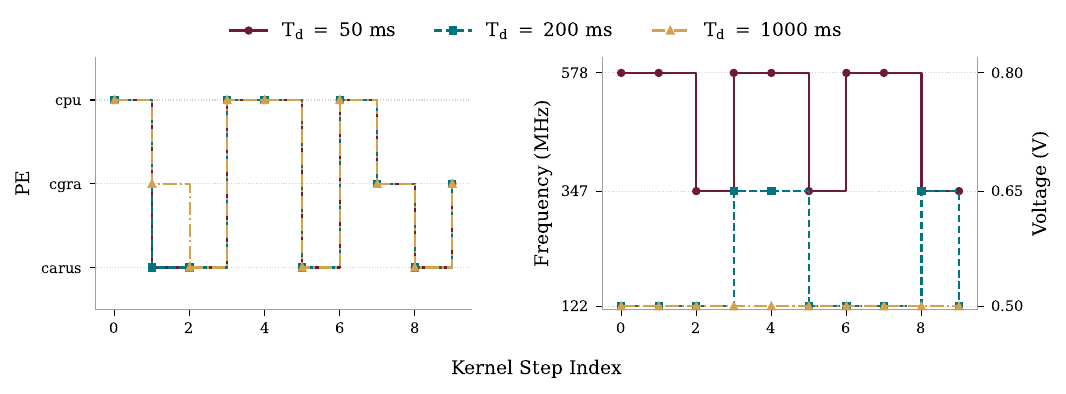} 
    \caption{Snapshot of {\schedname} scheduling decisions (PE allocation and V-F level per kernel) for a portion of the TSD transformer workload under timing constraint of \SI{1000}{\milli\second}, \SI{200}{\milli\second}, and \SI{50}{\milli\second}.} 
    \Description{Per-kernel schedule excerpts for three deadlines, showing how MEDEA changes processing-element assignments and voltage-frequency levels as the timing constraint tightens.}
    \label{fig:sched_diff_constraints}
\end{figure}

As established in Section~\ref{sec:manger:algo}, {\schedname} achieves the energy savings demonstrated above by solving an optimization problem that seeks to minimize active energy ($E_{t,a}$) under the given timing constraint ($T_d$). 
The core decisions from this optimization involve selecting the PE and V-F setting for each kernel, with the tiling mode being a pre-determined choice for each PE-VF pair. 
To visualize these decisions in action, Figure~\ref{fig:sched_diff_constraints} provides a snapshot of {\schedname}'s generated {decisions for an illustrative sequence of kernels from the TSD transformer model} under the three evaluated timing constraints.
The figure clearly illustrates that as the timing constraint becomes more stringent (e.g., tightening from $\SI{1000}{\milli\second}$ down to $\SI{50}{\milli\second}$), {\schedname} strategically utilizes 
higher V-F settings to ensure timeliness, a direct application of kernel-level DVFS.
Under the relaxed $\SI{1000}{\milli\second}$ deadline, {\schedname} operates PEs at the lowest available V-F setting (e.g., $\SI{0.5}{\volt}$), as this is sufficient to meet the deadline. 
In such scenarios where the kernel-level DVFS knob offers limited room for further optimization, the {marginal} energy improvement becomes less {pronounced}.

\begin{figure}[htbp] 
    \centering 
    \includegraphics{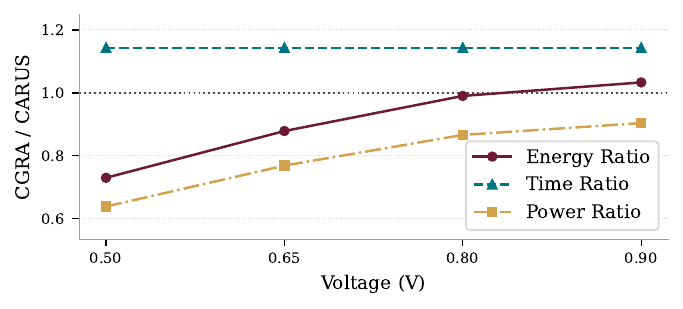} 
    \caption{TSD matmul subset: Ratio of CGRA performance and energy metrics (Energy, Power, Time) over Carus metrics versus V-F setting.}
    \Description{Line plot of CGRA-to-Carus energy, power, and execution-time ratios across HEEPTimize voltage-frequency settings, illustrating a voltage-dependent energy-efficiency crossover.}
    \label{fig:voltage_accel_comp}
\end{figure}

More interestingly, Figure~\ref{fig:sched_diff_constraints} reveals that the optimal choice of PE for a given kernel can also change with the deadline. While V-F adaptation is an expected strategy, this dynamic PE re-assignment highlights a more complex tradeoff. 
This occurs because the relative energy efficiency of the PEs (e.g., CGRA vs. Carus) is itself dependent on the operating V-F point.
To demonstrate this effect explicitly, we executed a representative TSD matmul subset (executable on both accelerators) on Carus and the CGRA individually across HEEPTimize's V-F range ($\SI{0.5}{\volt}$ to $\SI{0.9}{\volt}$). 
Figure~\ref{fig:voltage_accel_comp}, which presents the ratio of key performance and energy metrics (CGRA/Carus), shows that the relative average power consumption of the CGRA (to Carus) decreases significantly at lower V-F points. Given a constant operational cycle count ratio, this directly shifts their relative energy efficiency: 
the CGRA becomes comparatively more energy-efficient at lower voltages, while Carus tends to be more efficient at higher V-F settings. 
This observed efficiency crossover is characteristic of HULPs that integrate accelerators with disparate power profiles, such as the logic-dominant CGRA versus the more memory-influenced NMC where power scales differently with voltage. 
This behavior creates a complex optimization space and underscores the necessity for {\schedname} to dynamically co-optimize PE selection and V-F settings.

\subsection{Impact Analysis of Core {\schedname} Features} \label{sec:results:features_analysis}

To dissect the energy savings achieved by {\schedname} and {to} quantify the independent contribution of its core optimization strategies, we conducted a detailed feature impact analysis. 
{This evaluates the energy savings of {\schedname} while disabling one key feature at a time during the optimization process, while keeping all other mechanisms active.} 
This also helps illustrate the potential energy inefficiencies in related works that lack these capabilities. 
The core features analyzed are: (1) kernel-level runtime DVFS, (2) kernel-level scheduling, and (3) memory-aware adaptive tiling.

\begin{figure}[htbp]
    \centering
    \includegraphics[width=\linewidth]{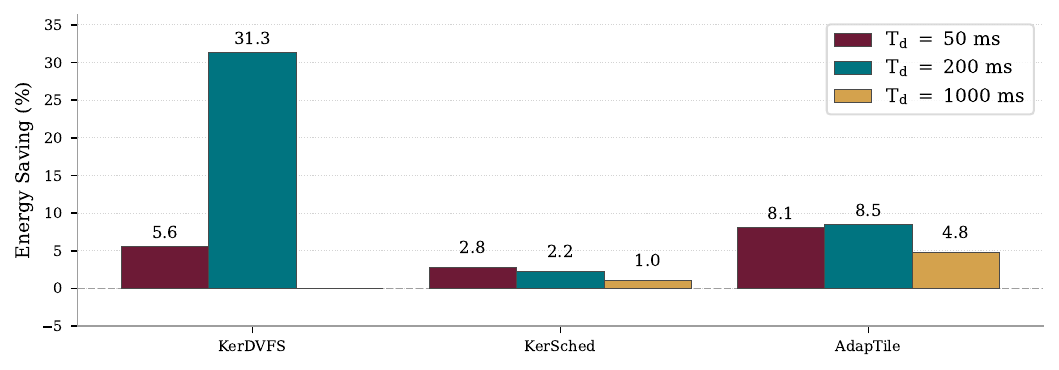} 
    \caption{{Energy saving when selectively disabling core {\schedname} features for the TSD model under different timing constraints.}}
    \Description{Bar chart of energy savings attributable to kernel-level DVFS, adaptive tiling, and kernel-level scheduling at 50, 200, and 1000 millisecond deadlines.}
    \label{fig:features_impact}
\end{figure}

Table~\ref{tab:feature_analysis} presents the absolute total energy consumption ($E_t$) for these configurations in the three timing constraints ($T_d$) for the TSD model. 
{To better visualize the impact, Figure~\ref{fig:features_impact} shows the corresponding percentage energy reduction achieved by each feature. 
This percentage of savings is calculated as $(E_{\text{w/oFeat}} - E_{\text{Full}}) / E_{\text{w/oFeat}} \times 100$, where $E_{\text{w/oFeat}}$ is the energy consumed with that specific feature disabled.}

\begin{table}[htbp] 
\centering
\caption{Energy consumption (\si{\micro\joule}) for {\schedname} feature analysis under different timing constraints.}
\label{tab:feature_analysis}
\begin{tabular}{l rrr} 
\toprule
{Sched. Setup} & \multicolumn{3}{c}{{Deadline (\si{\milli\second})}} \\
\cmidrule(lr){2-4} 
& {50} & {200} & {1000} \\
\midrule
Full {\schedname}        & 946  & 395   & 468   \\
\quad w/o KerDVFS             & 1002 & 576 & 468 \\
\quad w/o AdapTile               & 1030   &  432   & 492   \\
\quad w/o KerSched & 974 & 404  & 473  \\
\bottomrule
\end{tabular}
\end{table}

\subsubsection{Impact of Kernel-level DVFS}

{First, we analyze the impact of kernel-level DVFS.} 
We compare the full {\schedname} against a configuration where all other features are active, but a single application-level DVFS (AppDVFS) setting is used. 
{This AppDVFS setting is the lowest V-F point that allows the schedule to meet the deadline.} 
Figure~\ref{fig:features_impact} shows that {enabling} kernel-level DVFS (i.e., using AppDVFS) {yields no additional energy savings} for the $\SI{1000}{\milli\second}$ deadline. 
This is expected, as under such a relaxed constraint, {\schedname} {already uses the lowest V-F setting for all kernels}, making kernel-level DVFS adjustments redundant. 
However, for the $\SI{200}{\milli\second}$ deadline, {enabling} kernel-level DVFS {provides} a significant {$\SI{31.3}{\percent}$ energy saving}. 
For the $\SI{50}{\milli\second}$ deadline, the saving is $\SI{5.6}{\percent}$. 
This demonstrates that kernel-level DVFS is most impactful when the deadline allows for a diverse range of V-F settings across kernels; 
for very tight deadlines (such as $\SI{50}{\milli\second}$), most kernels are forced to higher V-F settings, reducing the scope for differential benefits of DVFS compared to an optimized AppDVFS.


\subsubsection{Impact of Kernel-level Scheduling}
Although kernel-level DVFS and scheduling are intertwined, we isolate the impact of scheduling granularity. 
Here, we compare the full {\schedname} approach against a configuration that uses coarse-grained scheduling (grouping kernels as defined in Section~\ref{sec:results:evaluation} for the CoarseGrain-AppDVFS baseline) 
but still applies group-level DVFS and all other optimizations. 
As seen in Figure~\ref{fig:features_impact}, {moving to fine-grained, per-kernel scheduling provides an energy saving between $\SI{1.0}{\percent}$ to $\SI{2.8}{\percent}$} across the deadlines. 
Although modest, this indicates that even with per-group DVFS, forcing a single PE/V-F choice for a diverse group of kernels can lead to suboptimal energy assignments for individual kernels within that group compared to per-kernel decision-making.

\subsubsection{Impact of Adaptive Tiling}
Finally, we assess the contribution of {\schedname}'s memory-aware adaptive tiling {by comparing it} against a configuration {with a fixed, non-adaptive tiling strategy (always defaulting to double-buffering, $t_{db}$, regardless of kernel size or memory pressure) while keeping all other features active}.
{Enabling} adaptive tiling, as shown in Figure~\ref{fig:features_impact}, {results in energy savings of $\SI{8.1}{\percent}$ ($\SI{50}{\milli\second}$), $\SI{8.5}{\percent}$ ($\SI{200}{\milli\second}$), and $\SI{4.8}{\percent}$ ($\SI{1000}{\milli\second}$).} 
The consistent {savings} highlight the importance of dynamically selecting the best tiling mode ($t_{sb}$ vs. $t_{db}$) based on kernel characteristics to optimize data movement and memory utilization. 
The slightly lower impact at $\SI{1000}{\milli\second}$ might be influenced by longer idle times, which could mask smaller differences in active energy due to tiling choices.

\subsection{\cmod{Design-Time Optimization Cost}} \label{sec:results:ilp_cost}

While the preceding results concern \emph{runtime} inference on HEEPTimize, {\schedname}'s MCKP/ILP solver runs once on a host during deployment preparation and is never invoked on the device during inference (Sec.~\ref{sec:manger:algo}).
The design-time problem size is therefore the relevant scalability metric: it is bounded by the number of kernels~$N_{\mathcal{W}}$, processing elements~$N_{\mathcal{P}}$, and V--F points~$|S_{vf}|$, \emph{not} by the number of DNN weight parameters or tensor elements.
Before the ILP is built, {\schedname} evaluates each candidate configuration~$\omega_{ij}=(p_{ij},v_{ij},c_{ij})$ and records its precomputed active time and energy; tiling modes are resolved beforehand, so they do not multiply the decision space (Sec.~\ref{sec:manger:algo}).
Table~\ref{tab:ilp_cost} reports the resulting problem size and wall-clock breakdown for the TSD workload; Sec.~\ref{sec:limitations:optimization} discusses why this offline cost is acceptable for HULP deployment.

\begin{table}[htbp]
    \centering
    \caption{\cmod{Design-time ILP/MCKP cost for the TSD workload on HEEPTimize (Intel Core i7-1068NG7 @ \SI{2.30}{\giga\hertz}, \SI{32}{\gibi\byte} RAM, macOS~26.2).}}
    \label{tab:ilp_cost}
    \small
    \begin{tabular}{lr}
        \toprule
        Quantity & Value \\
        \midrule
        Kernels $N_{\mathcal{W}}$ & 233 \\
        PEs $N_{\mathcal{P}}$ / V--F points $|S_{vf}|$ & 3 / 4 \\
        Avg.\ configs per kernel $|\Omega_i|$ (min--max) & 8.4 (4--12) \\
        Binary variables $\sum_i |\Omega_i|$ & 1968 \\
        Constraints ($N_{\mathcal{W}} + 1$) & 234 \\
        \bottomrule
    \end{tabular}

    \medskip
    \begin{tabular}{lrrrr}
        \toprule
        Deadline $T_d$ & $\Omega$ enum. & ILP build & CBC solve & End-to-end\\
        & (\si{\second}) & (\si{\second}) & (\si{\second}) & (\si{\second}) \\
        \midrule
        \SI{50}{\milli\second}  & 0.017 & 0.073 & 0.29 & 0.38 \\
        \SI{200}{\milli\second} & 0.017 & 0.057 & 1.45 & 1.53 \\
        \SI{1000}{\milli\second} & 0.017 & 0.038 & 0.08 & 0.13 \\
        \bottomrule
    \end{tabular}
\end{table}

CPU-only kernels (e.g., Softmax, GeLU) contribute only $|S_{vf}|$ configurations each.
Configuration enumeration and ILP construction together account for less than \SI{0.1}{\second} in all cases; CBC dominates only under moderately tight deadlines.
CBC time peaks at $T_d = \SI{200}{\milli\second}$ because the optimal schedule is near-binding ($T_{t,a} \approx T_d$), leaving many competing low-energy configurations for branch-and-bound to discriminate; at \SI{50}{\milli\second} most kernels are forced to high V--F (a narrower feasible set), and at \SI{1000}{\milli\second} ample slack ($T_{t,a} \approx \SI{223}{\milli\second}$) simplifies the search.
The worst-case solver time (\SI{1.45}{\second} at $T_d = \SI{200}{\milli\second}$) remains below the enforced inference windows.

Table~\ref{tab:ilp_scaling} synthetically grows the TSD workload by increasing the number of transformer encoder blocks (kernel shapes and HEEPTimize profiles are otherwise unchanged), at a fixed relaxed deadline $T_d = \SI{1000}{\milli\second}$.
$N_{\mathcal{W}}$ and binary variables increase approximately linearly ($540 \rightarrow 3872$ for 1--8 layers); $\Omega$~enumeration, ILP construction, and CBC solve time grow accordingly.
The four-layer row matches the \SI{1000}{\milli\second} entry in Table~\ref{tab:ilp_cost}; tighter-deadline CBC times for the evaluated model appear in the same table.

\begin{table}[!htbp]
    \centering
    \caption{Scaling sweep at $T_d = \SI{1000}{\milli\second}$: varying transformer encoder depth (same HEEPTimize platform).}
    \label{tab:ilp_scaling}
    \small
    \begin{tabular}{rrrrrr}
        \toprule
        Encoder layers & $N_{\mathcal{W}}$ & Binary variables & $\Omega$ enum.\ (\si{\second}) & ILP build (\si{\second}) & CBC solve (\si{\second}) \\
        \midrule
        1 & 65  & 540  & 0.003 & 0.010 & 0.03 \\
        2 & 121 & 1016 & 0.006 & 0.019 & 0.04 \\
        4 & 233 & 1968 & 0.017 & 0.038 & 0.08 \\
        6 & 345 & 2920 & 0.018 & 0.061 & 0.12 \\
        8 & 457 & 3872 & 0.023 & 0.078 & 0.15 \\
        \bottomrule
    \end{tabular}
\end{table}

\FloatBarrier

%% file: secs/limitations.tex

\begin{cmodb}
\label{sec:limitations:scope}
\label{sec:limitations:workload}
{\schedname} is a \emph{design-time} manager for a single latency-constrained DNN inference: one sequential kernel workload~$\mathcal{W}$ with one deadline~$T_d$ (Eqs.~\eqref{eq:workload_def}--\eqref{eq:constraint_timing_final}).
The MCKP/ILP solver runs once on a host during deployment preparation, and the device executes the resulting fixed per-kernel schedule at runtime without invoking the optimizer (Secs.~\ref{sec:manger:algo},~\ref{sec:intro}).
Although optimization is offline, the deployed schedule still applies per-kernel V--F settings during inference; what is not re-optimized at runtime is PE assignment, tiling mode, or operating point when workload structure or resource availability changes unexpectedly.
The formulation targets serialized, statically characterized pipelines: concurrent multi-DNN co-execution, shared-resource contention under overlapping traffic, mixed non-DNN background tasks, dynamic control flow, and burst task arrivals are outside the current model.
Timing and power costs are drawn from profiled, deterministic characterization~($S_c$, $S_P$).
Operator lowering and model--platform mapping are upstream of scheduling.
{\schedname} optimizes PE assignment, V--F level, and tiling over a given~$\mathcal{W}$; it does not select among mathematically equivalent operator implementations (e.g., native vs.\ approximated Softmax/GeLU).
In our evaluation, such mapping choices are applied uniformly to all baselines on the same kernel list (Sec.~\ref{sec:setup:app}), so reported gains isolate scheduling and energy-management decisions rather than operator substitution.

The workload is modeled as a static, sequentially ordered kernel list~$\mathcal{W}$ (Eq.~\eqref{eq:workload_def}): list index encodes data-dependency order for feed-forward inference pipelines, and kernels execute strictly one after another at runtime.
{\schedname} does not optimize execution order or model fork--join parallelism within a single inference.
Highly branched or dynamically controlled graphs (e.g., early exit, mixture-of-experts routing) require a DAG workload representation or per-path offline schedules and are left for future work.
The evaluation focuses on the feed-forward Transformer-based TSD workload (Sec.~\ref{sec:setup:app}).
At the framework level, {\schedname} is DNN-architecture agnostic: any network that lowers to typed kernels in~$\mathcal{T}_{ops}$ with characterized profiles can, in principle, be scheduled once platform data~$S_c$ and~$S_P$ exist (Sec.~\ref{sec:manger:inputs:app}).
We evaluate on a transformer because its heterogeneous operator mix is structurally richer than the CNN-centric workloads common in prior HULP mappers~\cite{hamdi2024match, burrello2021dory, stahl2023fused}, not because {\schedname} is attention-specific.
We do not empirically demonstrate generalization across network topologies: the experimental validation is intentionally centered on one deep case study---the sequential TSD transformer pipeline under three deadlines---rather than a broad multi-architecture sweep; this depth-first focus is typical when per-platform characterization is costly~\cite{hamdi2024match, burrello2021dory, stahl2023fused} (Sec.~\ref{sec:setup:app}).

\label{sec:limitations:platform}
\label{sec:limitations:dvfs}
{\schedname}'s platform model is not an idealized zero-overhead memory abstraction: it captures a realistic HULP memory hierarchy---off-chip flash for weights, a shared L2 buffer, and private PE local memories---with operands reaching accelerators only after explicit flash$\rightarrow$L2 and L2$\rightarrow$LM transfers, reflecting that PEs such as Carus (NMC) operate on private local memory rather than shared L2~\cite{caon2024scalable} (Secs.~\ref{sec:manger:inputs:plat_static},~\ref{sec:setup:platform}).
Timing profiles~$S_c$ measured on HEEPTimize FPGA prototypes include end-to-end kernel cycles with tiled execution and DMA traffic, not compute-only idealizations; the timing model~$\mathcal{G}_T(\cdot)$ further accounts for data-movement cycles according to the selected tiling mode ($t_{sb}$ vs.\ $t_{db}$).
The formulation does not model dynamic bus bandwidth contention or L2 interference among concurrent bus masters; this omission is consistent with our sequential single-pipeline execution scope, in which measured per-kernel profiles reflect isolated DMA costs rather than multi-master arbitration under concurrent multi-application traffic.

Active times and energies in the current evaluation use steady-state profiled values at each selected voltage--frequency point via~$\mathcal{G}_T(\cdot)$ and~$\mathcal{G}_P(\cdot)$ (Eqs.~\eqref{eq:kernel_time_def}--\eqref{eq:constraint_timing_final}); they do not add the extra delay or energy incurred when the supply voltage changes between consecutive kernels.
Off-chip voltage regulators typically introduce transition latencies on the order of tens to hundreds of microseconds, whereas integrated on-chip regulators can reduce this to nanoseconds~\cite{keller2017risc,zidar2024dynamic}---a concern when scheduled units are very short.
The MCKP structure already supports pair-dependent transition penalties without changing decision variables: for configuration~$\omega_{ij}$ selected at kernel~$k_i$,
\[
T_a'(\omega_{ij}) = \mathcal{G}_T(\omega_{ij}) + \delta\!\bigl(v_{i-1}, v_{ij}\bigr),
\qquad
E_a'(\omega_{ij}) = \mathcal{G}_P(\omega_{ij})\,\mathcal{G}_T(\omega_{ij}) + \varepsilon\!\bigl(v_{i-1}, v_{ij}\bigr),
\]
where $\delta(\cdot)$ and $\varepsilon(\cdot)$ are obtained from platform PMU/regulator characterization.
Platform-specific measurement of the HEEPTimize power-delivery path and a transition-aware re-evaluation of the TSD schedules are identified as future work.

\label{sec:limitations:optimization}
{\schedname}'s ILP/MCKP solver incurs zero on-device optimization overhead: compilation-time cost is paid once per $(\mathcal{W}, T_d, \text{platform})$ tuple on a host and does not affect inference deadlines on the HULP.
Problem size scales with the number of kernels~$N_{\mathcal{W}}$, processing elements~$N_{\mathcal{P}}$, and V--F points~$|S_{vf}|$---not with DNN weight count or tensor element cardinality.
Operational sizes~$s_i$ enter only as inputs to pre-characterized timing and power profiles; tiling modes are resolved before the ILP, so they do not multiply the decision space (Sec.~\ref{sec:manger:algo}).
With fixed kernel order, the formulation is a compact multiple-choice knapsack ($|\text{variables}| = \sum_i |\Omega_i| \le N_{\mathcal{W}} N_{\mathcal{P}} |S_{vf}|$, $|\text{constraints}| = N_{\mathcal{W}} + 1$), structurally lighter than graph-level ILP schedulers that search over parallel DAG embeddings with precedence, communication, and memory-coupling constraints.

Sec.~\ref{sec:results:ilp_cost} quantifies design-time cost for the TSD workload on HEEPTimize: Table~\ref{tab:ilp_cost} reports problem size and per-deadline wall-clock breakdown; Table~\ref{tab:ilp_scaling} sweeps encoder depth at fixed~$T_d$.
Worst-case CBC solve time is acceptable for offline deployment preparation (on-device exact optimization over joint kernel--DVFS--tiling decisions remains prohibitive on milliwatt-class edge nodes).

\label{sec:limitations:future}
Natural extensions include:
(i)~multi-DNN and multi-application scheduling with shared-resource contention models and per-application deadlines or priorities;
(ii)~mixed non-DNN workloads via reserved time or bandwidth budgets;
(iii)~runtime elasticity---profile-drift detection, burst-aware re-optimization, and lightweight on-device re-scheduling when workload structure changes; the compact MCKP structure could serve as a warm-started design-time re-optimizer triggered on drift or task arrival, keeping heavy optimization off the inference critical path;
(iv)~finer interconnect and bus contention modeling for concurrent DMA masters and accelerator traffic;
(v)~branched and DAG workloads (early exit, mixture-of-experts, fork--join parallelism) via path-dependent offline schedules or runtime path selection;
(vi)~hierarchical deployment in which {\schedname} produces per-model schedules coupled with an outer runtime allocator;
(vii)~automated platform profiling to reduce manual characterization effort;
(viii)~tighter integration with compilation and mapping front-ends (e.g., ConSmax, MATCH-style lowering) that determine operator implementations upstream of scheduling;
(ix)~transition-aware DVFS re-evaluation on HEEPTimize using measured $\delta(\cdot)$ and $\varepsilon(\cdot)$ weights.
\end{cmodb}

%% file: secs/conclusion.tex
This paper has addressed the critical challenge of achieving energy-efficient DNN inference on resource-constrained HULPs. 
We introduced {\schedname}, a novel design-time multi-objective manager that uniquely integrates kernel-level scheduling, kernel-level runtime DVFS, and memory-aware adaptive tiling, all driven by a 
timing-constrained energy optimization strategy. Through a comprehensive evaluation on HEEPTimize, a custom HULP, {\schedname} demonstrated its ability to deliver significant energy efficiency, 
achieving energy reductions of up to $\SI{38}{\percent}$ compared to the relevant adapted baselines while consistently meeting strict application deadlines. 
Our in-depth analyses further quantified the significant individual contributions of {\schedname}'s core features. 
\cmod{Overall, this work has proposed a robust and adaptive manager essential for realizing the full potential of HULP systems for complex AI workloads, paving the way for more sophisticated intelligent applications on severely constrained devices.
A concise account of {\schedname}'s scope, modeling assumptions, and future extensions is given in Sec.~\ref{sec:limitations}.}

%% file: secs/acknowledgments.tex
The authors would like to thank María José Belda for her help in providing the kernels for CGRA, Michele Caon for his help in providing the evaluated DNN kernels for NMC, and Simone Machetti, Davide Schiavone and Luigi Giuffrida for their support on the platform setup part.
We thank Jan Tomasz Juraszek for his help during a semester project on verifying that our TSD optimizations maintained acceptable model performance.
This research is carried out in the frame of the “UrbanTwin: An urban digital twin for climate action: Assessing policies and solutions for energy, water and infrastructure” project with the financial support of the ETH-Domain Joint Initiative program in the Strategic Area Energy, Climate and Sustainable Environment. Also, this work was supported in part by the Swiss State Secretariat for Education, Research, and Innovation (SERI) through the SwissChips research project; and partially supported by Grant PID2024-158251NA-C33 funded by MICIU/AEI/10.13039/501100011033 and by FEDER, UE.